\documentclass[12pt,a4paper]{article}
\usepackage[latin1]{inputenc}
\usepackage{amsmath}
\usepackage{amsfonts}
\usepackage{amssymb}

\usepackage{graphicx}

\newcommand{\R}{{\mathbb R}}

\author{Moritz Duembgen and L.~C.~G. Rogers}
\title{Estimate Nothing}
\begin{document}

\maketitle


\begin{abstract}
In the econometrics of financial time series, it is customary to take 
some parametric model for the data, and then estimate the parameters from 
historical data. This approach suffers from several problems. 
Firstly, how is estimation error to be quantified, and then taken into account 
when making statements about the future behaviour of the observed
time series? 
Secondly, decisions may be taken today committing to future actions
over some quite long horizon, as in the trading of derivatives;  if
the model is re-estimated at some intermediate time, our earlier decisions
would need to be revised - but the derivative has already been traded at
the earlier price.
Thirdly, the exact form of the parametric model to be used is generally
taken as given at the outset; other competitor models might possibly
work better in some circumstances, but the methodology does not allow
them to be factored into the inference.  
What we propose here is a
very simple (Bayesian) alternative approach to inference and action 
in financial econometrics which deals decisively with all 
these issues. The key feature is that nothing is 
being estimated.
\end{abstract}

\section{Introduction.}\label{intro}
Following the original paper of Black \& Scholes \cite{BlackScholes} on 
the pricing of European options, published in 1973, it was quickly 
realized that the predictions of the model did not fit observed prices
very well, and in the intervening forty years a small army of 
alternative models has been assembled to try to fit the implied 
volatility surface. Without trying to be exhaustive, there are models
based on alternative diffusion assumptions, such as the CEV model
\cite{Beckers};
there are models with stochastic volatility, such as \cite{Heston},
\cite{SteinStein};
 there are models
which allow discontinuities in the price process, such as 
\cite{MertonJumps}, \cite{MadanVG}, \cite{CGMY}, \cite{Meixner},
\cite{Kou}; 
Markovian regime-switching models, such as \cite{Buffington}; 
models based on GARCH
dynamics, such as \cite{Duan}, to mention just some of the approaches tried.
In all cases, what is proposed is some parametric model for the
dynamics of the underlying asset, which is generally required to be
sufficiently simple that computation of option prices can be performed
either in closed form, or otherwise by an efficient numerical procedure.
This requirement is inescapable, because the methodology used to fit
the model to historical data requires repeated calculation of derivative
prices for different parameter vectors in a numerical search for the
best-fitting parameter vector, by whatever criterion is chosen. A single
calibration exercise may then require the calculation of thousands of 
derivative prices, so each one has to be very fast.  Commonly, the
fitting criterion used is least-squares, which is simple to work with, and
under an assumption of Gaussian observational errors also maximum-likelihood.

Once the best-fitting parameter vector has been identified, what is
then done with it? What is commonly done in the industry is that this
particular parameter vector is fed into some derivative-pricing
formula, and some formula for a hedging delta, treating this {\em 
estimate} as if it were the known true value. Now everyone who does
this knows that this ignores estimation error, but they do it anyway.
It would be comforting to believe that practitioners doing this have
always quantified the error in the estimate in some way, and that 
this is accounted for in the answers they pass to their colleagues.
But it is not easy to deal with these errors. Do we try to state 
a confidence set for the parameter? In multi-dimensions, what shape would
this set have? How would we discover the possible range of derivative
prices as the parameter varied over this confidence set? What conclusions
should be drawn for hedging deltas? How much priority would such checks
be given in a commercial environment?

The approach we propose in this paper can be viewed as a form of 
particle filtering, or Sequential Monte Carlo as it is known in the 
engineering literature. The use of such methods in financial 
econometrics is not new, and some of the important contributions in 
this area are discussed in the following literature survey.

Johannes \& Polson \cite{johannes2003mcmc} give a survey of 
computational Bayesian methods and their use in econometrics.
Darsinos \& Satchell \cite{DarsinosSatchell} propose a Gamma-Gaussian
prior for the precision and mean of the asset returns, which they 
update according to Bayes' rule, and then use as a mixing distribution
over the Black-Scholes option price formula to arrive at option
prices. The paper of Guidolin \& Timmermann \cite{GT} carries through
a similar analysis in the context of a binomial model, where the 
probabilities are treated as Bayesian parameters.
In contrast, Jacquier \& Jarrow \cite{JJ} use the disagreements between
model prices and market prices as the object of the Bayesian analysis;
this leads to typically quite complicated likelihoods over
the parameter space which have to be explored by MCMC methodology. 
Polson \& Stroud \cite{PolsonStroud} incorporate both the likelihood
of transitions, and the likelihood from the difference between the
model market prices, using the Heston model as the test example,
and again applying MCMC methodology to probe the posterior likelihood.
The thesis \cite{Gupta} of Alok Gupta discusses the general method
and illustrates it with the example of local volatility models.
Avellaneda {\it et al} \cite{Avellaneda_etal} propose the use of a
Feed Forward Neural Net to model implied volatility.
 

Bunnin {\it et al.} \cite{BGR} use Bayesian methods to make a comparison
of the standard Black-Scholes model, and a CEV model, using 
FTSE data. This paper is closest in spirit and methods to the
present contribution, so it may be worth highlighting the differences.
As here, \cite{BGR} actually make a Bayesian comparison of (two)
{\em completely different} models for the underlying data, and this 
we believe is an important extension of the traditional particle
filtering thinking, where typically the parametric model would be 
fixed and the inference would be attempting to find out about the 
unknown parameter. It is important that we can not only use Bayesian
methods to make inference among the representatives of a particular
parametric class of models, but we can also use it to compare
 {\em between} classes of models. As we shall see, this is of value 
 because when we come to look at some actual data, we see that when 
we let a large number of popular models compete, then at some times
we find one model doing well (that is, having high posterior
likelihood) and at other times our beliefs move rather to other
model classes.  This is what we would hope and expect. We do not
expect that {\em one} particular model will outperform all others in 
all situations; we realize that different models may be more or less
suitable at different times and in different markets, and the Bayesian
approach allows us to blend consistently the virtues of many different
models.  Apart from some differences of a technical nature concerning 
the calculation of the transition densities of the underlying processes,
the major difference between \cite{BGR} and our approach is that we
include in the model the log-likelihood for the fitting error of the
model prices to market prices.   Thus  we find in the literature 
broadly two different approaches to inference; one attempts to 
fit models by least squares to option price data, but pays no 
attention to the observed moves of the underlying; and the other 
looks at the moves of the underlying to perform inference on that
process, then passes that information through the model pricing function
to make statements about option prices.  We do both; and we find 
that sometimes a model which does a very good job of fitting option
data is not very credible when we come to look at what it says about
moves of the underlying, and {\it vice versa.}

The present study has the following features:
\begin{itemize}
\item Models from completely different families are compared and
combined;
\item The likelihood accounts for the moves of the underlying
assets as well as the discrepancies between model and market 
prices;
\item The log-likelihood contribution for the discrepancies
between market and model prices is not just a simple sum of squares
of differences, implying that we consider the errors to be 
independent across all derivatives, but is allowed to have a more
general (and more appropriate) Gaussian structure.
\end{itemize}
Earlier studies have incorporated some of these features, but 
to our knowledge this is the first time that all have been included.

Computationally, a Bayesian approach can be quite cumbersome because
the log-likelihood function over the (usually high-dimensional)
parameter space is so complicated that only MCMC techniques can 
possibly be contemplated. Moreover, such approaches are not well
suited to an adaptive algorithm, in view of the time taken to 
compute. We use what is in effect a simple
form of particle filtering, so that at any time the universe of models
under consideration is finite and known. In our calculations, we have
fixed a suitably-chosen universe of models, and worked just with
those; more sophisticated variants of particle filtering could be 
applied to adapt the universe of models as new data comes in, but 
the approach we present is good enough to get started.  For a 
general survey of the particle filtering literature, see, for
example, \cite{cappe2007overview}.

The methodology is then applied to daily data on the S\&P500 index
and options on that index, gathered over the period
January 2006 to December 2012, seven years in total,
covering a very wide range of market conditions.
We compare a wide range of different models: Black-Scholes and CEV
as representatives of diffusion models; Heston, Bates, SABR as 
representatives of stochastic volatility models; and VG, NIG, Kou, Merton,
as log-L\'evy examples.  The data period studied starts before the 2008
crash, and covers that turbulent period and some years afterwards. 
Interestingly, we find that as the market evolves, the posterior
probability shifts around the different model types, sometimes
favouring one, at other times another.

The plan of the paper is as follows. In Section \ref{S2} we
set some notation and explain what we do. Most of this is extremely
simple, but the one point of methodological interest is the way
we  treat the log-likelihood contribution of the differences between
market and model implied volatility, which is not the obvious first
choice.  In Section \ref{S4}, we explain the choices
made in implementing the numerical scheme,
and present some results. We exploited the excellent
Premia package, which provided us with efficient pricing  code for 
most of the models in the study. 
The CEV and SABR models were not available, and had to be separately
coded.
The Premia package was called from within NSP.
In Section \ref{conc} we conclude, and discuss future directions
for research.

\section{Modelling set-up.}\label{S2}
The central object of study is a single asset, whose log-price
at time $t$ will be denoted $X_t$. This is observed in discrete
time only, at the times $t=h, 2h, \ldots$. Also observed at these
times will be the prices of $A$ derivative securities; the market
price  of derivative $a$ observed at time $t$ will be denoted $Y^a_t$.
The evolution of $X$ is assumed to be Markovian, but the exact 
transition mechanism is not known with certainty; we shall suppose
that there are $J$ possible models for this evolution, and model $j$
has transition density 
\begin{equation}
p_j(x,x')  = P_j(\;  X_h \in dx' \;| 
\; X_0 = x  \;)/ dx'\qquad (j = 1, \ldots, J).
\label{transdensj}
\end{equation}
 It should be stated 
immediately that this template does not fit the stochastic volatility 
examples included in the study. For these examples, the transition
density has to depend on the current level of volatility as well
as the current spot price. In such cases, we shall understand
$p_j$ as the density of $X_h$ conditional on the known values of 
$X_0$ and the volatility at time $0$, and we may even approximate this
transition density (which is typically quite hard to obtain in 
closed form) by assuming that the volatility remains constant
over $[0,h]$. If $h$ is quite short - one day in our data - 
this assumption is reasonable, and is preferable to getting dragged
into some clumsy and slow computation. Anyone who finds these
assumptions objectionable is of course free to leave all stochastic
volatility models out of the comparison.

Associated to model $j$ is a pricing function $\varphi_j$
which returns a vector $(\varphi^a_j(X))_{a = 1}^A$ of
prices which depend on the current spot log-price $X$.  
Building the code to instantiate these pricing functions is of 
course a non-trivial task, but, with the availability of the Premia
software, one which we can consider done, even if it remains quite
challenging to wrap the Premia routines in the code which does 
the Bayesian model comparison.   The log-likelihood $\ell_j(t)$
of model $j$ at time $t$ is defined by $\ell_j(0) = 0$, and the
recursion
\begin{equation}
\ell_j(t) = \ell_j(t-h) + \log p_j(X_{t-h},X_t) 
-Q ( \varphi_j(X_t), Y_t),
\label{ell_1}
\end{equation}
where $Q$ is a non-negative definite quadratic form defined on 
$\R^A \times \R^A$.  The most obvious choice would be to take
$Q(y,z) \propto \Vert y-z \Vert^2$, a multiple of the squared
Euclidean distance, but as we shall argue below, this is not the 
best choice for the applications we have in mind.  The
interpretation of the quadratic term in \eqref{ell_1} is of course
that the difference between the market observed prices and 
the model prices is supposed to be some Gaussian random vector.
In practice, 
what we shall do is to generalize the recursion \eqref{ell_1} 
slightly to become
\begin{equation}
\ell_j(t) = \beta\, \ell_j(t-h) + \log p_j(X_{t-h},X_t) 
-Q ( \varphi_j(X_t), Y_t),
\label{ell_2}
\end{equation}
where $\beta \in (0,1]$ allows for some `forgetting' of the past.
This is a rather rough operational variation of the basic Bayesian
story, which can be justified in a simple linear Gaussian situation
by allowing assumed constant parameters to evolve as Gaussian 
random walks - see \cite{Muth}.  Though this is not the situation
we find ourselves in, some gradual downweighting of historical 
likelihood contributions permits more recent data to count more
heavily in our inference, and this we maintain is a natural
flexibility to request\footnote{In
practice, we just fixed a sensible value for $\beta$
and left it alone for all the 
calculations.
}. The fastidious reader can simply assume
that $\beta =1$ throughout.

At this point, we simply resort to Bayes' rule: at time $t$, 
the posterior distribution $\pi(t)$ over the $J$ models is just
given by
\begin{equation}
\pi_j(t) \propto \exp( \; \ell_j(t) \; ).
\label{Bayes}
\end{equation}
It is important to realize that at this point {\em everything}
becomes easy:
\begin{itemize}
\item If you want to know the distribution of $X_{t+h}$, it is 
given by the density 
\[
\sum_j \pi_j(t)\,  p_j(X_t, \cdot);
\]
\item If you want to give price for some exotic derivative, 
use model $j$ to calculate the price $\xi_j$ of the derivative, 
and take $\bar\xi \equiv \sum_j \pi_j(t) \, \xi_j$ as the price;
\item If you want to know how reliable the price $\bar\xi$ for
the exotic derivative is, you have a distribution over possible
prices, assigning weight $\pi_j(t)$ to value $\xi_j$, and from 
this you can assess the likely range of variation of price;
\item If you want to delta-hedge some derivative, you just
calculate the hedge $\theta_j$ which model $j$ would tell you to 
take, and then put on the position $\sum_j \pi_j(t) \, \theta_j$.
\end{itemize}
It is even more important to realize that {\em nothing has been
estimated!} What we have done is to take a fixed finite universe
of $J$ models, and we have {\em calculated} the posterior 
distribution over that universe of models - the numerical values
$\pi_j(t)$ are not estimates, they are true calculated values.

\subsection{Choice of $Q$.}\label{choiceQ}
Despite this simplicity, there are critical choices to be made
in the analysis, and the most important of these is the choice of
the quadratic form $Q$.  At this point, we will specify that the
derivatives under consideration are European call options, with 
expiries 1m, 2m, 3m, 6m, 1y, 1.5y, 2y, and strikes of moneyness
80\%, 90\%, 95\%, 100\%, 105\%, 110\%, and 120\%. Thus on each 
day of the dataset, we have the prices\footnote{In fact,
we have the implied volatilities, and we have to make use of the
riskless rates of USD interest to convert to prices.} of 49
instruments.  If we now take the quadratic form $Q$ to be the 
natural first choice
\begin{equation}
  Q(y,z) = \sum_a \frac{ |y_a - z_a|^2
  }{w^2_a}
  \label{choice_1}
\end{equation}
for weights $w_a$ chosen to be of the same magnitude as the 
bid-ask spread, we find that the numerical values are so large
as to completely swamp the contributions to the log-likelihood 
due to the transition functions. This appears initially to be 
an obstacle, but if we think what the form \eqref{choice_1}
is saying, we see how to get around it. What \eqref{choice_1}
says is that the observed values $Y^a_t$ are the model
values $\varphi_j^a(X_t)$ plus independent Gaussian errors.
But is this consistent with what we would expect to see?  If this
story held true, then the observed values would form a very rough 
surface, but we know from no-arbitrage considerations that the
market price surface {\em must} be convex in strike and increasing
in expiry.  So we do not expect this probabilistic model to 
explain the differences between the surfaces well. 

Another thought experiment which shows that this likelihood
cannot be a good choice is to consider what would happen if we were
to be given option prices of more and more strikes and expiries; 
we would see a surface convex in strike and increasing in expiry
filling in before our eyes.  We should expect that as we fill in 
more and more of this surface, the likelihood penalty should 
converge to some finite limit, which will not happen for 
\eqref{choice_1}.

So what would be a better choice?  To understand this, let
$C(\tau,K)$ (respectively, $C^{(j)}(\tau,K)$)
denote the 
market (respectively, model-$j$) 
 call price of expiry $\tau>0$ and strike
$K$;  for now, let us suppose that
spot has been scaled to 1, and the riskless rate is zero.
We know that $C$ is convex in $K$, decreasing to zero, and 
that $C_K(\tau,0) = -E S_\tau = -1$. Thus $-C_K(\tau, 
\cdot)$ can be interpreted as the tail of a distribution
function. What we therefore propose 
to take as the quadratic penalty $Q$  for model $j$ is 
\begin{equation}
Q( C^{(j)},C) = \lambda \int_0^T \int_0^\infty\{ \, C_K(v,K)
-C_K^{(j)}(v,K)  \, \}^2 \; \; dK\,dv
\label{Qdef1}
\end{equation}
for some $\lambda >0$.  Since $0 \leq C_K \leq 1$ always, easy
estimation gives us that $Q(C^{(j)},C) \leq 2\lambda T$,
which is finite.   In the general situation, we shall 
suppose that the spot value $S_0 = \exp(X_0)$ gets scaled out
by defining the renormalized call price
\begin{equation}
c(\tau,k) = C(\tau,kS_0)/S_0,
\label{cdef}
\end{equation}
with the analogous definition
\begin{equation}
 Q( c^{(j)},c) = \lambda \int_0^T \int_0^\infty\{ \, c_k(v,k)
-c_k^{(j)}(v,k)  \, \}^2 \; \; dk\,dv
\label{Qdef2}
\end{equation}
of the quadratic penalty.

How does this work out when it comes to the finite data set we
are given?  Thus we know model and market prices for 
moneyness values $0< k_1< k_2 < \ldots < k_N$, and expiries
$0 < \tau_1 < \ldots < \tau_M$, and need to approximate $Q$
as given by \eqref{Qdef2}. We shall approximate the time
integral by the trapezium rule, allowing us to concentrate
just on the inner integral over moneyness for some fixed value
$\tau$ (say) of expiry. Define $k_0 = 0$, and set 
$z_j = c(\tau, k_j)$, with the obvious definition $z_0 = 1$, the 
value of a zero-strike call.  Then we have\footnote{We 
treat the model call prices $c^{(j)}(\tau,\cdot)$ analogously.}
\begin{equation}
c_k(\tau,x) = \frac{z_j - z_{j-1}}{k_j - k_{j-1}}
\qquad   \hbox{\rm for $k_{j-1}<x\leq k_j$}
\label{cdisc}
\end{equation}
and $c_k(\tau,x) = 0$ for $x > k_N$.  The inner integral over
moneyness thus reduces to a finite sum which is easy to evaluate.

\medskip
There remains the issue of choosing the scaling $\lambda$ in the
definition of $Q$, and here it is less clear how one should proceed.
Operationally, the key requirement is to set $\lambda$ at a value
such that the contribution to the log-likelihood \eqref{ell_1}
from the moves of the underlying and from the call surface fitting
errors {\em should be of similar orders of magnitude}; we do not
want to find that we are only fitting the call surface, or only
fitting the asset dynamics. There is no unique recipe here;
we could take a training data set, calculate the two types of 
contributions for all the models we are studying, and then fix
$\lambda$ so as to equalize the log-likelihood contributions from
the two components. Alternatively, we could use the training dataset
just for the Black-Scholes models as a way of choosing $\lambda$.
Or again, we could run the calculations for a few different values 
of $\lambda$ to see how the results differ. The freedom to choose
here corresponds exactly to the freedom to choose a model for
the observational errors;
 we postulate that the call surface fitting errors have
a particular Gaussian structure, but we need to make a choice 
about the scaling. Such modelling choices are universal in statistics.

\section{Methodology and results.}\label{S4}

It is a general rule of thumb that in order to do a good job 
matching the prices of call options, a model with at least four
parameters will be needed: one for the centring, one for the 
variance, one for the right tail, one for the left tail.
The models we took (with the number of parameters in parentheses)
were:
Black-Scholes (1); CEV (2); Heston (4); SABR (3);
 Bates (7); Merton (4); Kou (4);
Variance-Gamma (3); and Normal Inverse Gaussian (3).

For each of
these models, we have to have ways to calculate the transition
density and the option prices very efficiently, and in 
Appendix \ref{app} we record closed-form formulas - where 
available  - or computational approaches otherwise, for all
of these models.  We reduced the dimensionality of the Bates model 
by assuming that 
$\mu^J = \sigma^J$.

To make the approach work, we need to pick a suitably representative
family of models to put into the comparison, and this is perhaps the
most methodologically challenging aspect. There are two general
approaches:
\begin{itemize}
\item[(i)] Adaptively select the models in the population;
\item[(ii)] Choose a set of models at the start and never
change them.
\end{itemize}
It is easier to deal with the second at a conceptual level, 
but computationally this can easily become prohibitive. To focus the
discussion, suppose that we have some parametric model whose
parameter space is $\R^d$. Without any initial information, we might
try to cover some interval in $\R^d$ by setting up a rectangular
grid; but if we were to ask for just ten grid points in each 
coordinate, we would end up with $10^d$ parameter vectors. This 
is perfectly feasible for  $d = 1, 2$, but for $d=3$ is getting
moderately large, $d=4$ is becoming rather uncomfortable, and $d\geq 5$ 
is best avoided - and this is for a very coarse grid. Other ways
of placing the points in $\R^d$ could be used; for example, we could
set down some randomly-generated set of points, or we might use
some pseudo-random sequence, or perhaps a `spread out' set of
points such as those developed\footnote{Files
of these points can be downloaded from {\tt http://www.quantize.maths-fi.com/}.
} by Pag\`es and Printems \cite{PP}.  However, none of these 
methods actually avoid what is perhaps the main pitfall, which is 
that it is perfectly possible that {\em all} of the parameter
vectors chosen are very bad choices. Without knowing what `typical'
values should be, it would only be too easy to put down a grid in 
a box which was very distant from what are `typical' parameter
choices - and this issue becomes more pressing the larger $d$.
There simply is no alternative to carrying out some initial 
search to locate some region where the parameter values are 
reasonable, and the simplest way to do this is to do some 
least-squares fitting of the model to data.

How should we do this? In principle, we could take a set of 
training data, perhaps using option price data taken on the first
day of the month for each of the previous 20 months, and then 
do a least-squares fit of the model to that data. This would then 
give us 20 least-squares-fitted parameter vectors which do 
fit hopefully quite well on at least one day, and we could use these
vectors, and perhaps other scattered around them, to create our
fixed set of parameter vectors. This works well up to a point, 
but with the data that we used, we found that as we moved through
the seven-year time period, most of the models we had chosen at
the start got more and more unlikely {\it a posteriori}.  The reason
for this was very simple - the world was very different in 
January 2008 from 
how it had been in January 2006.  Model parameters that made 
sense in January 2006 were no longer at all relevant.  This made
it essential that the models used were informed by what was
happening throughout the period of study. The honest way to do 
this is to use the first approach, which is to adaptively select
models as time passes, but we did not do this as it is quite 
difficult to build an algorithm which does this 
effectively.  The point is that we have to 
devise a mechanism to allow the individual particles to move 
around as new information comes in, and the moves must not just
be random - we have to do some importance sampling, moving the
new particles towards `more likely' parameter regions.
And how are we to identify `more likely' regions? By least-squares
of course.  But now the adaptive approach requires us to do 
many least-squares optimizations per time period, so is in danger
of running very, very slowly.  Of course, we can just do occasional
least-squares calculations to apply an occasional steer to 
the particle population, but in the end for any real problem, 
any SMC approach has to incorporate something akin to a least-squares
optimization, which is heavy on computational time.  The calculations
can readily be parallelized, but the computational challenge is 
substantial however it is to be done. For actual application, the
new data is coming in quite slowly, once a day, so if it takes
1000s compute time to calculate the updating of the particle
population, this is not an issue; but it is an issue for backtesting
where we want to evaluate the algorithm on several years of 
historical data.

What we actually did involved an element of data-snooping. 
We selected a small number of days spread out
through the sample where we
did a least-squares fit of the model to implied volatility
and return histograms.  This gave us a number of parameter
vectors which were reasonable for each model for the data
under study.  This is of course cheating, but it
allows us to assess the performance of the method assuming that 
we are able to build a particle filter that adaptively varies
the particle population to follow the region where the likelihood
is currently concentrated.  In more detail, what we did was:
\begin{itemize}
	\item[(i)] calibrated each model to volatility surfaces as 
	well as histograms at several points in time, throughout our sample; 
	\item[(ii)] For each parameter, we then span a rough grid of around
	 $5-10$ points between the highest and lowest value it ever took 
	 during these calibrations;
	\item[(iii)] This resulted in a multidimensional grid of up to 
	several thousand parameter combinations for each model;
	\item[(iv)] We then ran the Bayesian analysis 
	for each model's parameter combinations singly;
	\item[(v)] In order to bring the number of parameter combinations 
	per model down to $100$, we then selected only those parameter
	 combinations which were at some point the ``best'' of their 
	 class, or within a certain distance to the best of their class 
	 (where the distance was chosen such that $100$ parameter 
	 combinations resulted).
\end{itemize}

\begin{figure}
\centering
\includegraphics[scale=0.75,angle=0]{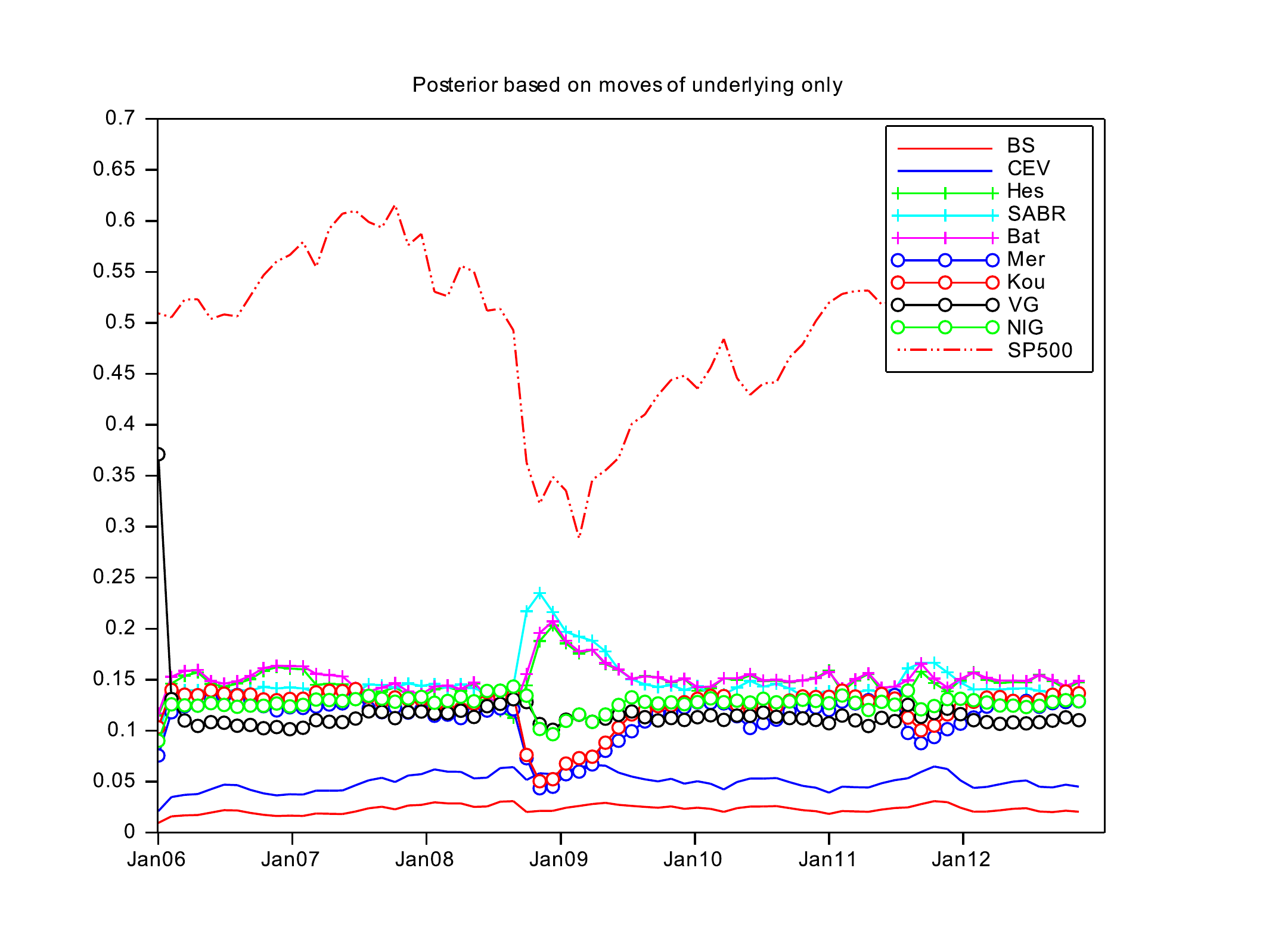}
\caption{Posterior from  moves of the underlying only.}
\label{Fig1}
\end{figure}

\begin{figure}
\centering
\includegraphics[scale=0.75,angle=0]{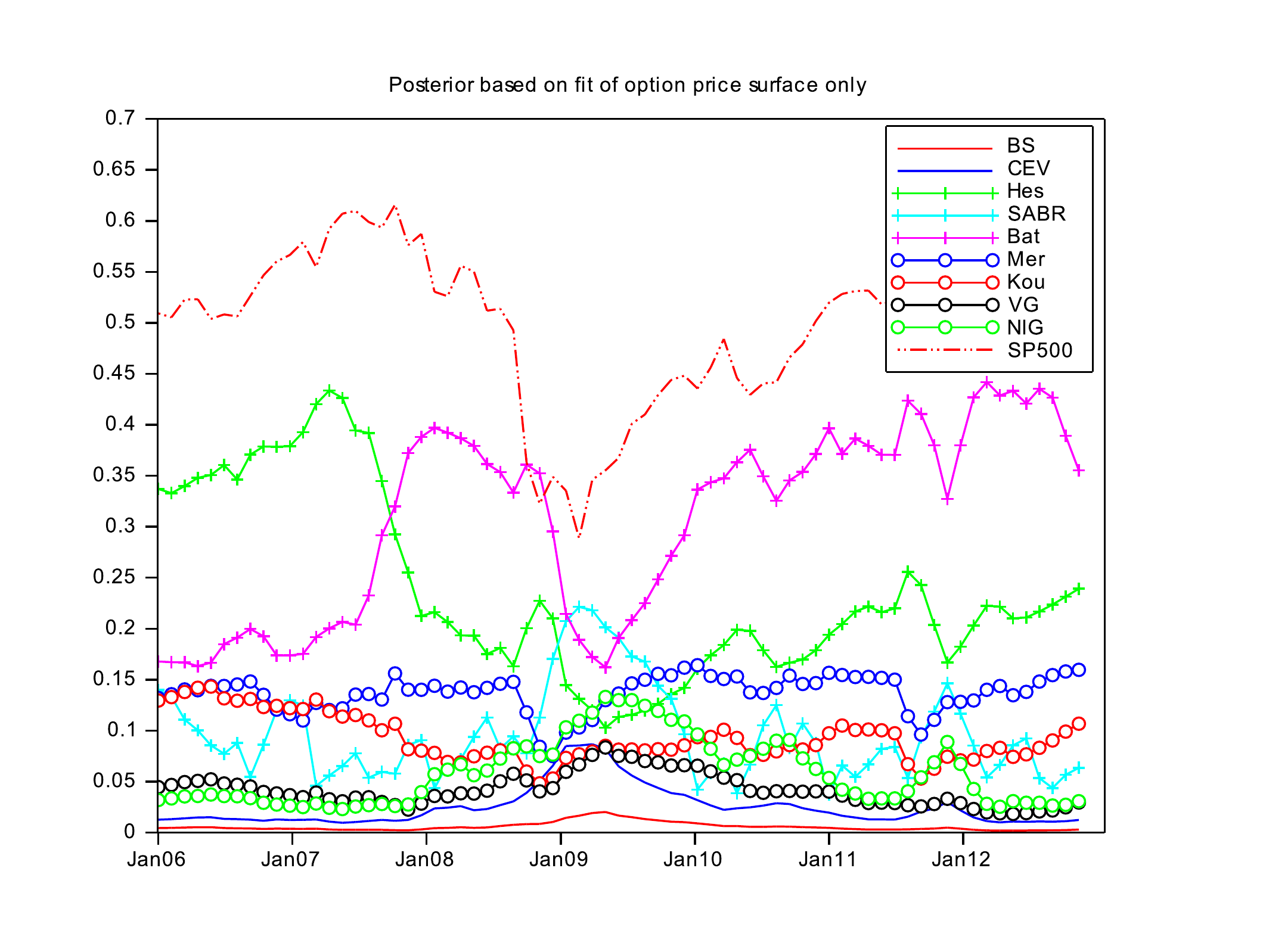}
\caption{Posterior from options prices only.}
\label{Fig2}
\end{figure}

\begin{figure}
\centering
\includegraphics[scale=0.75,angle=0]{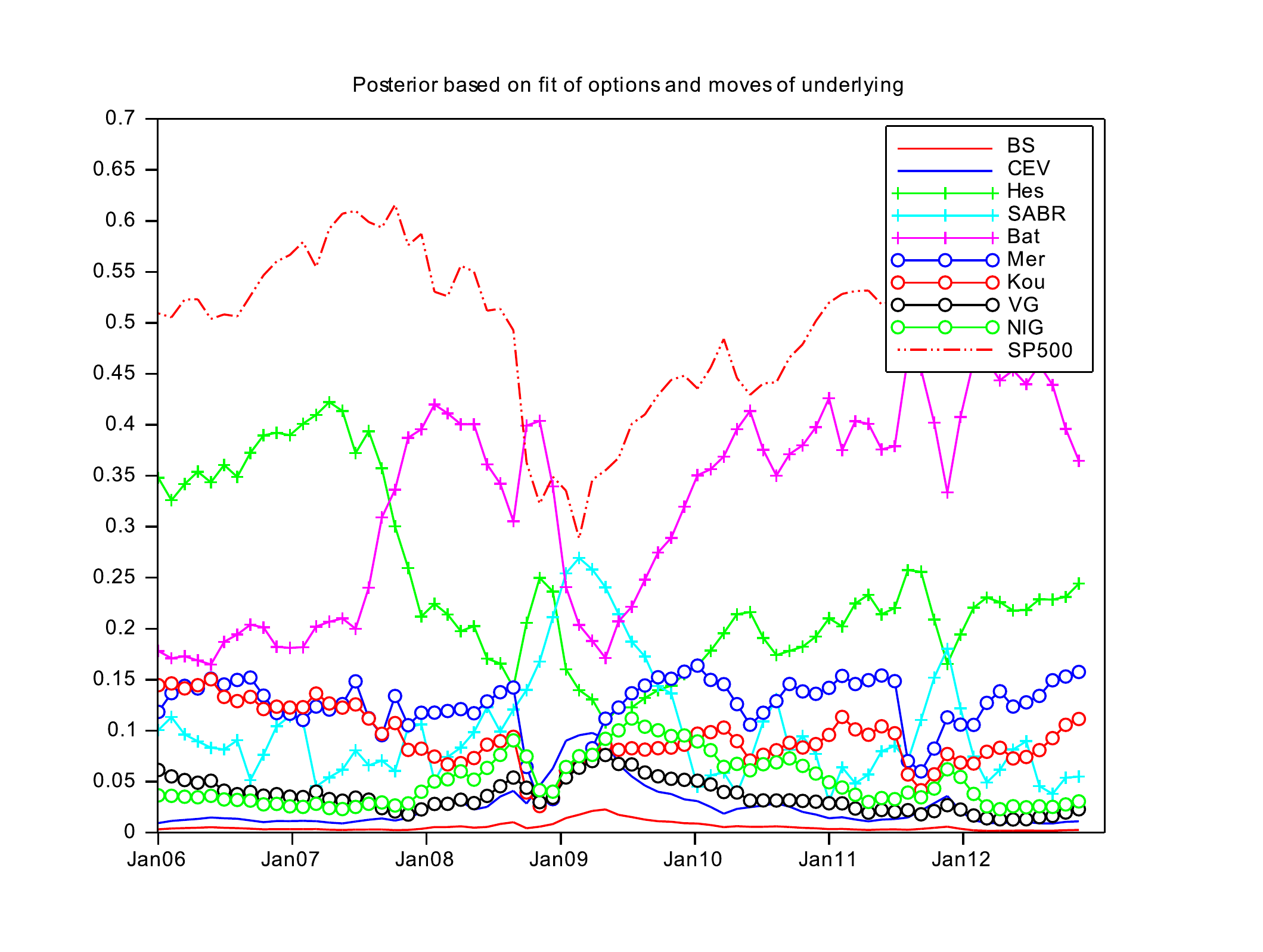}
\caption{Posterior from moves of the underlying and fit of option prices.}
\label{Fig3}
\end{figure}

The results are displayed in Figures \ref{Fig1}, \ref{Fig2} and \ref{Fig3}.
Each figure contains ten plots; all show in dashed the realized path of the 
S\&P500 index\footnote{.. suitably scaled to fit into the range of the
posterior probabilities ..} over the time period, 
to anchor the plots of posterior weights
to what was happening globally. There are then nine further plots, two for 
the diffusion models, Black-Scholes and CEV; three for the stochastic volatility
models, Heston, SABR and Bates; and four for the log-L\'evy models Merton, Kou, 
variance-gamma and normal inverse Gaussian. Each of the models was represented
in the Bayesian comparison by hundreds of instances, each instance corresponding to 
a different choice of the parameter vector; and the posterior plot for that model 
was obtained by adding up the posterior probabilities for all the instances of
that model.  

When we compare based only on the likelihood contribution coming from the 
transition densities of the underlying S\&P500 index, Figure \ref{Fig1}, what
we see is:
\begin{itemize}
\item The diffusion models Black-Scholes and CEV do poorly, with Black-Scholes
worse than CEV. Since Black-Scholes is a special case of CEV, this is not
unexpected;
\item The remaining models perform comparably, except during the crisis period 
September 2008-2009, when the log-L\'evy models do noticeably worse than 
the stochastic volatility models. Again, this is not surprising, because this 
was a period of increased volatility, which the stochastic volatility models 
can accommodate, but the IID returns assumption common to all the log-L\'evy
models is not compatible with such a period of heightened volatility.
\end{itemize}

When we turn to the comparison (Figure \ref{Fig2}) based on the fit of the 
option price surface, things look rather different:
\begin{itemize}
\item Once again, the diffusion models Black-Scholes and CEV do poorly, 
with Black-Scholes worse than CEV;
\item The log-L\'evy models are now quite spread out, with Merton generally
doing best, variance-gamma generally doing worst, and Kou and NIG switching
second and third places in the ranking;
\item The stochastic volatility models do best, with SABR performing well 
during the crisis, while Bates and Heston do well outside the crisis period.
\end{itemize}

The final comparison, Figure \ref{Fig3}, looks qualitatively similar to 
Figure \ref{Fig2}; in view of the fact that just looking at moves of the 
underlying did not seem to separate the different non-diffusion models, 
it is perhaps not surprising that adding in the moves of the likelihood to the
comparison changes things only a little.

\section{Conclusions.}\label{conc}
This study has proposed a Bayesian modelling paradigm for inference on 
assets and derivatives written on those assets which is at once both old and
new. It is old in that the Bayesian principle has been known for 250 
years\footnote{Bayes' paper {\em  An Essay towards solving a Problem in 
the Doctrine of Chances}  was read to the Royal Society in 1763, two years
after his death. The paper presented there had been prepared for publication by 
the Welsh nonconformist preacher Richard Price.}, and there is nothing 
conceptually here but a  systematic application of the Bayesian
principle. Where there is novelty, it is in the use of both the moves of
the underlying asset and the market prices of options to derive the likelihoods;
and in the comparison of models of completely different types all at once.

Though the approach is very simple, the dividends are not insignificant.
Firstly and most importantly, the procedure permits a {\em completely
consistent} approach to model inference, in contrast to the estimation-based
`calibration' paradigm which is in almost universal use in the industry 
at the time of writing. There is never any doubt about what we should be doing 
to hedge or to mark-to-market a portfolio of derivatives, and whatever we do today
will be consistent with what we did before, and with what we will do in the future.
The big distinction is that calibration attempts to estimate, and then uses 
the estimates as if they were known true values - ignoring all estimation error.
In the physical sciences it may be a reasonable approximation to ignore estimation
error in many situations, but in financial markets
 the signal-to-noise ratio is almost  always tiny, and estimation error just
 cannot be ignored. There is a pointer to this in Figures \ref{Fig1}, \ref{Fig2},
 \ref{Fig3}; all the models considered, though very different, remain 
 competitive - none of them drop to insignificant posterior probability. Thus
 the data does not allow us decisively to reject any of them, 
 not even the over-simplified Black-Scholes model, despite having
 several years of data on very liquid assets.
 
 The second dividend is that it allows comparison of models of very 
 different types; we do {\em not} have to insist at the outset that the model
 for the data {\em has to be} Heston; or variance-gamma; or SABR. Each of the
 popular models has good and bad points, and the Bayesian methodology allows 
 for the combination of the strengths of all of them, while honestly treating
 the very obvious fact that we never know which (if any) of the proposed 
 models is correct. 
 
  As is to be expected, the extent to which such a 
 Bayesian analysis will succeed will depend heavily on the range of models
 put into the comparison, just as a frequentist analysis depends on the 
 assumed model; and this is inevitably a subjective choice. In a more advanced
 version of the story told here, a sophisticated adaptive model selection will
 be required, but this is a delicate art. It seems in general that the 
 particle-filtering methodology can work well in dimension up to about ten, but
 beyond that it requires a lot of careful adaptation to the special features of
 the problem, as is to be expected of a simple general method. Getting this 
 right requires a lot of effort, and is beyond the scope of this study.
 Nevertheless, the preliminary results reported here suggest that this effort
 will be worthwhile.

\pagebreak

\pagebreak
\bibliography{EN}
\bibliographystyle{plain}

\end{document}